\documentclass[pra,aps,amsfonts]{revtex4}
\def\textbf#1{{\bf #1}}
\def\>{\rangle}
\def\<{\langle}
\def\beq{\begin{equation}}

\def\eeq{\end{equation}}
\def\be{\begin{equation}}
\def\ee{\end{equation}}
\def\ben{\begin{eqnarray}}
\def\een{\end{eqnarray}}
\def\beqa{\begin{eqnarray}}
\def\eeqa{\end{eqnarray}}
\def\eea{\end{array}}
\def\bea{\begin{array}}
\newcommand{\bei}{\begin{itemize}}
\newcommand{\eei}{\end{itemize}}
\newcommand{\bee}{\begin{enumerate}}
\newcommand{\eee}{\end{enumerate}}
\newcommand{\tr}{{\rm tr}}

\def\blacksquare{\vrule height 4pt width 3pt depth2pt}

\def\hcal{{\cal H}}
\def\dcal{{\cal D}}
\def\scal{{\cal S}}
\def\mcal{{\cal M}}
\def\tr{{\rm Tr}}

\def\>{\rangle}
\def\<{\langle}

\def\ot{\otimes}

\def\fproof{\proof{f}}
\def\fmroof{\mroof{f}}
\def\fcond{F}
\def\farrow{f_\downarrow}
\def\farrowcpl{f_\downarrow^{cpl}}

\def\proof{\widehat}
\def\mroof#1{\mathop{#1}\limits^{\frown} }

\newtheorem{lemma}{Lemma}

\newtheorem{proposition}{Proposition}
\newtheorem{definition}{Definition}

\begin{document}

\title{On asymptotic continuity of functions of quantum states}
\author{Barbara Synak-Radtke$^{(1)}$ and Micha\l{} Horodecki$^{(1)}$ }

\affiliation{$^{(1)}$Institute of Theoretical Physics and Astrophysics,
University of Gda\'nsk, Poland}

\begin{abstract}
A useful kind of continuity of quantum states functions in asymptotic 
regime is  so-called asymptotic continuity.
In this paper we provide general tools for checking 
if a function possesses this property. 
First we prove  equivalence of asymptotic continuity with 
so-called {\it robustness under admixture}. 
This allows us to show that relative entropy distance from
a convex set including  maximally mixed state is asymptotically continuous. 
Subsequently, we consider {\it arrowing} - a way of building a new function 
out of a given one. The procedure originates from constructions of intrinsic information 
and entanglement of formation.  We show that arrowing preserves 
asymptotic continuity for a class of functions (so-called subextensive ones). 
The result is illustrated by means of  several examples. 
\end{abstract}

\maketitle

\section{Introduction}
One of basic issues of Quantum Information Theory is to evaluate operational quantities 
such as capacities of quantum (usual of teleportation) channel 
\cite{BennettDS97-cap,BennettDSS-cap2004} 
costs creating quantum states  under some 
natural constraints \cite{BDSW1996,cost}, compression rates 
\cite{Schumacher1995} or localisable information rates 
\cite{OHHH2001,huge-delta}. The quantities are usually defined in 
spirit of Shannon --  in asymptotic regime of 
many uses of channel or many copies of state. Apart from such operational
quantities one also considers mathematical functions, that are expected to reflect 
somehow those features of states or channels. To this end, one chooses functions, 
that satisfy some requirements. For example, most of entanglement measures 
are mathematical functions, that do not increase under local operations 
and classical communication \cite{BDSW1996,Vidal-mon2000}. Other examples 
are correlation measures 
(see e.g. \cite{HendersonVedral,IBMHor2002,DevetakW03-common,SynakH04-deltacl}).
Such functions turn out to be very useful, as they often provide upper 
or lower bounds for operational quantities. in asymptotic regime, 
the functions are especially useful, if they are {\it asymptotically continuous}.
The prototype for asymptotic continuity is Fannes inequality \cite{Fannes1973}
for von Neumann entropy $S(\varrho)=-\tr \varrho\log \varrho$.
which says that for any  states $\varrho$ and $\sigma$ with $||\varrho-\sigma||_1\leq 1/2$ 
we have 
\be
|S(\varrho)-S(\rho)|\leq ||\varrho-\sigma||_1 \log d +\eta(||\varrho-\sigma||_1)
\label{eq:Fannes}
\ee
where $\eta(x)=-x\log x$, $d$ is dimension of Hilbert space.
The important feature of this stronger form of continuity, is that the right-hand-side 
scales logarithmically with dimension of Hilbert space. 
This kind of inequality, was first applied in quantum information theory 
in \cite{MH-compression,BarnumCFJS-compr2000} to provide lower bound 
for compression rates of mixed signal states (interestingly, 
the question of achievability of the bound is in general still 
open). Subsequently, it was applied to entanglement theory \cite{limits}
which lead, in particular, to methods of providing bounds for 
distillable entanglement and entanglement cost \cite{DonaldHR2001,cost}.
Asymptotic continuity has become an important tool in proving irreversibility 
of pure states transformations (see \cite{Michal2001} and references therein).

In \cite{Nielsen-cont,DonaldH1999} two measures of entanglement 
have been proven to satisfy Fannes-like inequality (i.e. to be asymptotically 
continuous) -- entanglement of formation $E_F$ \cite{BDSW1996} and relative 
entropy of entanglement \cite{PlenioVedral1998}. In \cite{Alicki-Fannes} 
asymptotic continuity of conditional entropy $S(A|B)= S(\varrho_{AB})-S(\varrho_B)$ 
have been proven, where the right-hand-side depend only on dimension of system $A$. 
This allowed to prove asymptotic continuity of third measure of entanglement 
-- squashed entanglement \cite{Winter-squashed-ent}.  
The importance of asymptotic continuity was made even more transparent in \cite{LockQ}
where it was shown, that a convex and so called {\it subextensive} 
function, if not asymptotically continuous, it behaves in a quite weird 
way: namely, after removing one qubit, it can change at arbitrarily large  amount.

Clearly, it is very important to know whether a function is asymptotically continuous 
or not.  Yet it is usually rather a difficult task.  The aim of this paper 
is to provide general tools for checking asymptotic continuity. First, we show that 
the latter is equivalent to so called "robustness under admixtures",
i.e. a function is asymptotically continuous, if it does not change too much 
under admixing any state with a small weight. Using it, 
we prove that relative entropy distance from any convex set 
including maximally mixed state is asymptotically continuous, 
extending therefore result of \cite{DonaldH1999} where 
it was proven for compact and convex sets.

Next, we consider a procedure, called {\it arrowing}, of building new functions 
out of given functions. The procedure originates both  from 
classical privacy theory \cite{intI,Christandl:2002} -- where the prototype 
was so-called intrinsic information -- as well 
as from entanglement theory, since it includes as a special case 
the other procedure called {\it convex roof} \cite{Uhlmann-roof},
the prototype of which was entanglement of formation \cite{BDSW1996}.
Since arrowing is commonly used in different contexts (see quite recent application \cite{Dong}), it is important 
to be able to check the properties of arrowed versions of different functions. 
We provide here a quite general result, showing that for subextensive functions 
such procedure  preserve asymptotic continuity, i.e. if an original function 
is asymptotically continuous, so is its "arrowed" version.
We then apply it to show, that some tripartite 
entanglement measure \cite{LPSW1999,Michal2001} as well as 
so called {\it mixed convex roof} of quantum mutual information 
introduced in \cite{Christandl:2002} are asymptotically 
continuous.

\section{Basic definitions \label{sec:def}}
In this section we will introduce some definitions  which we will use throughout this paper.
{\it Set of states}. 
A positive operator $\varrho \in \mathcal{S}$ with $\tr\varrho=1$, acting on 
Hilbert space $\mathcal{H}$ we will call state. Set of all states will
 be denoted by $\scal(\hcal)$. (We will deal with finite dimensional Hilbert spaces).
A state is called pure, if it is of the form $|\psi \>\< \psi |$ where $\psi\in \hcal$. 
Otherwise it is called mixed state. \\
{\it Von Neumann entropy} $S(\varrho)$  for a state $\varrho $ is given by formula:
\ben
S(\varrho )=-\tr \varrho \log \varrho 
\een 
We use base 2 logarithm  in this paper. \\
{\it Relative entropy }  for 
states $\varrho $ and $\sigma $ is defined as:
\be
S( \varrho|\sigma )=\tr \varrho  \log \varrho  -\tr\varrho  \log \sigma 
\ee
{\it Trace norm } of an operator A is given by:
\be
||A||_1=\tr\sqrt{AA^\dagger}
\ee 
where $A^\dagger$ stands for Hermitian conjugation.\\
{\it Measurement.} We will consider measurements with finite number of 
outcomes, represented by finite sets of operators $\mcal=\{A_i\}$
satisfying $\sum_i A_i^\dagger A_i=I$. Slightly abusing terminology, we will 
call the measurements POVMs (Positive Operator Valued Measure).

{\it Subextensivity} A function $f:\scal(\hcal)\to R$ is subextensive if 
\ben 
&&\forall_{\varrho }\quad \exists_M \quad f(\varrho) \leq M \log d 
\een
where $M$ is constant, $d=\dim\hcal$.

\begin{definition}
Let $f$ be a real-valued function $f:S(\mathcal{C}^d )\mapsto \mathcal{R}$ and $\varrho_1$, $\varrho_2$ 
are states acting on Hilbert space $\mathcal{C}^d$ and $\varepsilon =||\varrho_1- \varrho_2 ||_1$.
Then  a function is asymptotically continuous if fulfills the following condition
\ben
\forall_{\varrho_1,\varrho_2}  |f(\varrho_1)-f(\varrho_2) |\leq K_1 \varepsilon \log d+O(\varepsilon )
\label{1}
\een
 where $K_1$ is  constant
 and $O(\varepsilon )$ is  any function, which satisfies the condition that $O(\varepsilon )$ converges to 0
when $\varepsilon $ converges to 0 and depends only on $\varepsilon $. 
(In  particular, it does not depend on dimension).
\end{definition}

\begin{definition}
Let $f$ be a real-valued function $f:S(\mathcal{C}^d )\mapsto \mathcal{R}$ and $\varrho_1$, $\varrho_2$ 
are states acting on Hilbert space $\mathcal{C}^d$. Then  a function is robust  under admixtures if
\ben
\forall_{\varrho_1,\varrho_2} \forall_{\delta>0} | f((1-\delta ) \varrho_1 + \delta  \varrho _2 )-f(\varrho_1 )|
\leq K_2 \delta \log d+O(\delta )
\label{2}
\een
 where $K_2$ is  constant
 and $O(\delta  )$ is  any function, which satisfies the condition that $O(\delta  )$ converges to 0
when $\delta $ converges to 0 and depends only on $\delta $. (In particular, it does not depend on dimension).
\end{definition}

 \textbf{Remark.}  Notice that usually for asymptotic continuity or robustness under admixtures we will not require fulfilling 
conditions (\ref{1}) and (\ref{2}) 
for whole range of $\varepsilon $ or $\delta $. 
We will rather restrict to some limited subset of positive real value of   $\varepsilon $ or $\delta $ 
(limited by 1 or $\frac{1}{2}$, for example.)

\section{Asymptotic continuity  and robustness under small admixtures.}

In this section we prove equivalence between asymptotic continuity
and robustness under admixtures of function. This is  an extension of result 
of \cite{LockQ}, where it is  proved that 
if a function $f$, under admixtures does not change more than a constant, 
and  subextensive then is also asymptotically  continuous.

\begin{proposition}
Let $f$ be a function $f:S(\mathcal{C}^d )\mapsto \mathcal{R}$ 
then the function is asymptotically continuous {\it if only if} is 
robust under admixtures.
\label{pr1}
\end{proposition}
\textbf{Remark.} This proposition can be also proved when we do not require "Lipschitz type" continuities, but rather "Cauchy type" ones.
(See appendix)\\


\textbf{Proof.}\\
$" \Rightarrow "$\\
We  assume that function is asymptotically continuous. This implies
\ben
&&|f((1-\delta ) \varrho_1+\delta  \varrho _2) -f(\varrho_1) |
 \leq K_1 ||\varrho_1-((1-\delta  ) \varrho_1+\delta  \varrho _2)||_1\log d+O(||\varrho_1-((1-\delta  ) \varrho_1+\delta  \varrho _2)||_1 )=\\
&&=K_1 ||\delta \varrho _1-\delta   \varrho_2||_1\log d+O(||\delta \varrho _1-\delta   \varrho_2||_1 )
\leq 2K_1 \delta \log d+O(2\delta )
\een
Lets take $K_2=2K_1$. Then 
\be
|f((1-\delta) \varrho_1+\delta \varrho _2) -f(\varrho_1) |\leq K_2\delta  \log d+O(\delta  )
\ee

$" \Rightarrow "$\\
We will base on result of Refs. \cite{Alicki-Fannes} (see also \cite{ArakiMor}), which can be viewed as a 
sort of generalized Tales theorem
\ben
\forall_{\varrho_1,\varrho_2} \quad 
\exists_{\sigma, \gamma_1 \gamma_2} \quad\sigma =(1- \varepsilon)\varrho _1+\varepsilon \gamma _1
=(1- \varepsilon)\varrho _2+\varepsilon \gamma _2  
\een
where $\varrho_1,\varrho_2,\sigma, \gamma_1 ,\gamma_2 $ are states acting on Hilbert 
space and $\varepsilon =||\varrho_1- \varrho_2||_1$ .
Using it we obtain:
\ben
&&|f(\varrho _2)-f(\varrho _1)|\leq |f(\varrho _2)-f(\sigma )|+|f(\sigma)-f(\varrho _1)|=\\
&&|f((1- \varepsilon)\varrho _2+\varepsilon \gamma _2  )-f(\varrho _2)|+
|f((1- \varepsilon)\varrho _1+\varepsilon \gamma _1)-f(\varrho _1)|
\leq 2K_2 \varepsilon \log d +2 O(\varepsilon ) 
\een
so that  we can take $K_1=2K_2$.
Then

\ben
|f(\varrho _2)-f(\varrho _1)|\leq K_1 \varepsilon  \log d + O(\varepsilon )
\een
This ends the proof.



\subsection{Application: asymptotic continuity of relative entropy distance 
from convex set of states.}

In \cite{DonaldH1999} it was shown that so called relative entropy 
distance from convex, compact set including maximally state $\frac{I}{d}$ is asymptotically continuous.
The proof was quit involved. Here, basing on Proposition \ref{pr1} we present a more general result, where we do not require 
compactness of the set. Moreover our proof is  straighter.

Relative entropy of distance $E_R^\mathcal{D}$ is
defined as follows
\be 
E_R^\mathcal{D}(\varrho)=\inf_{\sigma  \in \mathcal{D}}S( \varrho|\sigma )
\label{E_R}
\ee 
where  $\mathcal{D}$ is a convex set of state  including maximally mixed state, $\varrho \in\mathcal{C}^d$. 

 We start with lemma:
\begin{lemma}
Relative entropy of distance $E_R^\mathcal{D}$ fulfills the following condition
\be
|E_R^\mathcal{D}((1-\varepsilon) \varrho+\varepsilon\sigma )-E_R^\mathcal{D}(\varrho) |\leq 2\varepsilon \log d +H(\varepsilon) 
\ee
where $H(\varepsilon)=-\varepsilon \log\varepsilon -(1-\varepsilon )\log(1-\varepsilon )  $
\label{lem1}
\end{lemma}
\textbf{Proof.}\\
First we show the that $E_R^\mathcal{D}$ satisfies the following inequality
\be
\sum_k p_k E_R^\mathcal{D}(\varrho _k)-E_R^\mathcal{D}(\sum_k p_k\varrho _k)\leq S(\sum_k p_k\varrho _k)-\sum_k p_kS(\varrho _k)
\ee
This fact was shown for relative entropy distance from separable states in \cite{revers}, but it is also true for 
relative entropy distance from any convex set of states. Here we repeat this proof  for $E_R^\mathcal{D}$ defined in (\ref{E_R}).
Notice that for $\varrho=\sum_kp_k \varrho_k $
\ben
&&S(\varrho| \sigma) =S(\sum_k p_k\varrho_k| \sigma)=\tr(\sum_k p_k\varrho_k\log(\sum_k p_k\varrho_k)
-\sum_k p_k\varrho_k \log \sigma )=\\
&&=\tr(\sum_kp_k(\varrho _k\log\varrho _k-\varrho _k\log\sigma +\varrho _k\log\varrho-\varrho _k\log\varrho _k))=\\
&&=\sum_kp_kS(\varrho _k|\sigma )+\sum_kp_kS(\varrho _k)-S(\varrho )
\een
Let $\sigma \in \dcal$ be a state such that 
$E_R^\mathcal{D}=S(\varrho| \sigma)-\delta $.
Then we can rewrite
\ben
E_R(\varrho)=\sum_kp_kS(\varrho _k|\sigma )+\sum_kp_kS(\varrho _k)-S(\varrho )-\delta \geq 
\sum_kp_kE_R(\varrho _k)+\sum_kp_kS(\varrho _k)-S(\varrho )-\delta
\een
Since by definition of $E_R^\dcal$  $\delta$ can be arbitrarily small, we obtain
\be
\sum_k p_k E_R(\varrho _k)-E_R(\sum_k p_k\varrho _k)\leq S(\sum_k p_k\varrho _k)-\sum_k p_kS(\varrho _k)
\ee
We use also fact that \cite{W78}
\be
S(\sum_k p_k\varrho _k)\leq \sum_k p_kS(\varrho _k)+H(\{p_k\} )
\ee
and that relative entropy distance is convex function, what is implied by convexity of quantum relative entropy in two arguments.
Notice  also that $E_R$ is bounded by $\log d$, because $\mathcal{D}$ includes maximally mixed state 
(so $E_R \leq S(\varrho|\frac{I}{d})=\log d-S(\varrho )\leq \log d $).
Then we have
\ben
&&| E_R((1-\varepsilon) \varrho+\varepsilon\sigma )-E_R(\varrho)|
=| E_R((1-\varepsilon) \varrho+\varepsilon\sigma )-(1-\varepsilon)E_R(\varrho)-\varepsilon E_R(\sigma )
 -\varepsilon E_R(\varrho)+\varepsilon E_R(\sigma )|\leq\\
&&=| E_R((1-\varepsilon) \varrho+\varepsilon\sigma )-(1-\varepsilon)E_R(\varrho)-\varepsilon E_R(\sigma )|
+\varepsilon |E_R(\varrho)|+\varepsilon |E_R(\sigma )|\\
&&=(1-\varepsilon)E_R(\varrho)+\varepsilon E_R(\sigma )-E_R((1-\varepsilon) \varrho+\varepsilon\sigma )
+\varepsilon |E_R(\varrho)|+\varepsilon |E_R(\sigma )|\\
&&\leq S((1-\varepsilon) \varrho+\varepsilon\sigma )-(1-\varepsilon)S(\varrho)-\varepsilon S(\sigma ) +\varepsilon \log d+\varepsilon \log d
\leq  H(\varepsilon)+2\varepsilon \log d  
\een
This 
ends the proof.\\
\textbf{Remark.}
Note that the main feature of $E_R^\mathcal{D}$ responsible for robustness under admixtures, are the following:\\
1) $E_R^\mathcal{D}$ satisfy inequality :
\be
|E_R(\sum_kp_k\varrho_k)-\sum_kp_kE_R(\varrho_k)|\leq H(\{p_k\} )
\ee
2) $E_R^\mathcal{D}$ is  bounded by $\log d$.

\begin{lemma}
Relative entropy of distance $E_R$ is asymptotic continuous i.e.
\be
|E_R(\varrho)-E_R(\sigma ) |\leq 4 \varepsilon \log d+2H(\varepsilon )
\label{er}
\ee
where $H(\varepsilon)=-\varepsilon \log\varepsilon -(1-\varepsilon )\log(1-\varepsilon )  $ and $\varepsilon=||\varrho- \sigma ||_1$.
\end{lemma}
\textbf{Proof.}\\
$E_R$ is robust under admixtures so under Proposition \ref {pr1} is also asymptotically continuous.

\section{Asymptotic continuity of functions built by "arrowing"}
\label{sec:arrow}

In this section we consider "arrowing" -- a construction that from given function  
$f$ creates a new function denoted by $f_\downarrow$. The definition is motivated 
by intrinsic information and its generalizations 
\cite{MaurerWolf00CK,Lutkenhaus,ChrRen04}.  The new function 
$f_\downarrow$ is  defined on  enlarged system
as follows
\begin{definition}
\label{def:arrow}
For any function $f:\scal(\hcal_X)\to R$ acting on states of system $X$,
we define function $\farrow:\scal(\hcal_X \ot \hcal_E) \to R$ 
as follows
\be
\farrow(\rho_{XE})=\inf_{\{A_i\}} \sum_i p_i f(\rho^i_X)
\ee
where infimum is taken over all finite POVM's  $\{A_i\}$  performed on system $E$
and 
\be
p_i=\tr (I_X \ot A_i) \rho_{XE},\quad \rho^i_X= {1\over p_i} 
\tr_E (I_X \ot A_i\,  \rho_{XE}\, I_X \ot A_i^\dagger )
\ee
i.e. $p_i$ is probability of outcome $i$, and $\rho^i_X$ is the state 
of system $X$ given outcome $i$ was obtained. 
\end{definition}
\textbf{Remark.} We can  define modified version of previous function as follows:  
\begin{definition}
For any function $f:\scal(\hcal_X)\to R$ acting on states of system $X$,
we define function $f_\uparrow:\scal(\hcal_X \ot \hcal_E) \to R$ 
as follows
\be
f_\uparrow (\rho_{XE})=\sup_{\{A_i\}} \sum_i p_i f(\rho^i_X)
\ee
where supremum is taken over all finite POVM's  $\{A_i\}$  performed on system $E$
and 
\be
p_i=\tr (I_X \ot A_i) \rho_{XE},\quad \rho^i_X= {1\over p_i} 
\tr_E (I_X \ot A_i\,  \rho_{XE}\, I_X \ot A_i^\dagger )
\ee
i.e. $p_i$ is probability of outcome $i$, and $\rho^i_X$ is the state 
of system $X$ given outcome $i$ was obtained. 
\end{definition}
All features of $f_\downarrow $ presenting in this paper are also valid for function $f_\uparrow $.
\vspace{1cm}

We have the following lemma, which is proven in Sec. \ref{ach}:
\begin{lemma}
The infimum in the definition of $\farrow$ is achievable. 
\end{lemma}

We will show in this section, that that asymptotic continuity and subextensivity of  
function $f$ implies asymptotic continuity of $f_\downarrow$. Thus in a sense, arrowing 
preserves asymptotic continuity. Let us stress that all the 
involved systems are finite-dimensional. 

We will need the following definition:
\begin{definition}
Given a function $f$ defined on states of a system $X$, 
we define its  conditional version $\fcond$ for a quantum-classical state of a system $XE$ 
\be
\rho_{XE}^{qc}=\sum_i p_i \rho^i_X \ot |i\>_E\<i|
\label{eq:qc-states}
\ee
as follows
\be
\fcond(\rho_{XE}^{qc})=\sum_ip_i f(\rho_X^i)
\ee
If the quantum classical state was obtained from state $\rho_{XE}$ 
by a POVM $\mcal$ performed on system $E$ we will also use notation
$\fcond(\rho_{XE},\mcal)\equiv\fcond(\rho_{XE}^{qc})$.
\end{definition}

Let us now present  the main result of this section.
\begin{proposition}
\label{prop:arrow}
Let  $f$ be a  function defined on states of system $X$,  which is 
subextensive and asymptotic  continuous. Then function $f_\downarrow $ is also 
asymptotically continuous. Moreover, the constant in asymptotic continuity 
condition depends only on dimension of system $X$. 
\end{proposition}
\textbf{Proof.}\\
Let  $\varrho_{XE}$ and $\sigma_{XE}$ be states and 
$\varepsilon=||\varrho_{XE}-\sigma_{XE}||_1$.  Let $\mcal_\varrho=\{A_k^\varrho\}$ 
and $\mcal_\sigma=\{A_k^\sigma\}$ 
be the optimal measurements for $\rho$ and $\sigma$ respectively (i.e. the ones 
achieving infimum in definition of $\farrow$) 
where $\sum_k A_k^\varrho{}^\dagger A_k^\varrho=I_E$, 
$\sum_k A_k^\sigma{}^\dagger A_k^\sigma=I_E$.
For measurement $\mcal_\sigma$, let $p_k$ and $q_k$ be probabilities of 
outcomes if a state was $\varrho$ and $\sigma$ respectively.
The resulting states on system $X$, given the outcome was $k$,  
 we will denote by $\varrho_k$ and $\sigma_k$ respectively. 
Due to asymptotic continuity (see sec. \ref{sec:def})
we assume that 
\be
|f(\varrho_{XE})-f(\sigma_{XE})|\leq K \varepsilon \log d_X + O(\varepsilon)
\ee
and due to subextensivity
\be
|f(\rho)|\leq  M \log d_X
\ee
for any state $\rho$ on system $X$, where $d_X=\dim \hcal_X$ and 
$M$ and $K$ are constants. 
Then we have the following estimate
\ben
&&f_\downarrow(\varrho_{XE} )-f_\downarrow(\sigma_{XE} )=\fcond(\varrho_{XE},\mcal_\varrho )
-\fcond(\sigma _{XE},\mcal_\sigma )\leq \fcond(\varrho_{XE},\mcal_\sigma  )-
\fcond(\sigma_{XE},\mcal_\sigma )=\\
&&=\sum_k p_k f(\varrho^k_{X})-\sum_k q_k f(\sigma ^k_{X})
\leq\bigl|\sum_k p_k f(\varrho^k_{X})-\sum_k q_k f(\sigma ^k_{X})\bigr|=\\
&&=\bigl|\sum_k p_k f(\varrho^k_{X})- p_k f(\sigma_{X}^k)+p_k f(\sigma^k_{X})- q_k f(\sigma^k_{X})\bigr|\leq \\
&&\leq \sum_k \bigl(p_k |f(\varrho^k_{X})- f(\sigma ^k_{X})|+|p_k - q_k|\, | f(\sigma^k_{X})|\bigr)\leq \\
&&\leq \sum_k p_k \varepsilon_k K\log d_{X}+\varepsilon M\log d_{X} +O(\varepsilon)
\leq K_1 \varepsilon \log d_{X}+O(\varepsilon )
\een
where  $\varepsilon_k=||\varrho^k_{X}-\sigma^k_{X}||_1$ and $K_1=2K+M$.
The last two steps of the above estimate are implied by asymptotic continuity,
subextensivity of the function $f$ and the following facts
(see \cite{Winter-squashed-ent}):
\be
\sum_k |p_k-q_k|\leq \varepsilon
\label{eq:fact1}
\ee
and 
\be
\sum_k p_k\varepsilon_k\leq 2\varepsilon 
\label{eq:fact2}
\ee
The inequality (\ref{eq:fact1}) we get via the following estimate
\ben
&&\sum_k |p_k-q_k|=\bigl\|\sum_k p_k |k\>\< k|-\sum_k q_k|k\>\< k|\bigr\|_1 \leq
\bigl\|\sum_k p_k \varrho^k_{X}\otimes| k\>\< k|-\sum_k q_k\sigma_{X}^k
\otimes|k\>\< k|\bigr\|_1=\\
&&||(I_X\otimes\Lambda_\sigma){\varrho}_{XE}-
(I_X\otimes\Lambda_\sigma){\sigma }_{XE}||_1
\leq || \varrho_{XE}-\sigma_{XE} ||_1=\varepsilon.
\een
where $\Lambda_\sigma$ is  a completely positive map
induced by POVM $\mcal_\sigma$ as follows
\be
\Lambda_\sigma(\cdot)=\sum_k \tr [A_k^\sigma (\cdot) A_k^\sigma{}^\dagger]  \,|k\>\<k|
\ee

We have used here the fact that 
trace norm does not increase under completely positive trace preserving maps 
\cite{Ruskai-tr}.

The inequality (\ref{eq:fact2}) is proven as follows
\ben
&&\varepsilon =||\varrho_{XE}-\sigma_{XE} ||_1\geq
\sum_k||p_k \varrho^k_{X}\otimes|k\>\<k|- q_k\sigma_{X}^k\otimes |k\>\<k|||_1=
\sum_k\|p_k \varrho^k_{X}- q_k\sigma_{X}^k\|_1 \\
&&\geq\sum_k(||p_k \varrho^k_{X}-p_k\sigma_{X}^k||_1-||p_k\sigma_{X}^k- q_k\sigma_{X}^k||_1)\\
&&=\sum_kp_k ||\varrho^k_{X}-\sigma_{X}^k||_1-\sum_k|p_k- q_k|)
\geq \sum_kp_k\varepsilon _k-\varepsilon 
\een
Analogously we can show that
\ben
&&f_\downarrow({\sigma}_{XE} )-f_\downarrow({\varrho}_{XE} )=F({\sigma}_{XE},\mcal_\sigma  )
-F({\varrho}_{XE},\mcal_\varrho )\leq F({\sigma }_{XE},\mcal_\varrho )
-F({\varrho  }_{XE},\mcal_\varrho )
\leq K_1 \varepsilon \log d_{X}+O(\varepsilon )
\een
Thus we obtain 
\ben
|f_\downarrow(\varrho^{XE} )-f_\downarrow({\sigma}^{XE} |\leq 
K_1 \varepsilon \log d_{X}+O(\varepsilon)
\een
This ends the proof. \blacksquare

{\bf Remark.} In the proof we have used the fact that the infimum in definition 
of $\farrow$ is achievable. However it is not essential:
the proof that does not use it is very similar to the above one. 

Finally, consider modification of the function $\farrow$,
where we do not optimize over all POVM's, but only over complete POVM's,
for which the operators $A_k$ are of rank one. 

\begin{definition}
For any function $f:\scal(\hcal_X)\to R$ acting on states of system $X$,
we define function $\farrowcpl:\scal(\hcal_X \ot \hcal_E) \to R$ 
as follows
\be
\farrow^{cpl}(\rho_{XE})=\inf_{\{A_i\}} \sum_i p_i f(\rho^i_X)
\ee
where infimum is taken over all finite POVM's  $\{A_i\}$ with elements 
$A_i$ being of rank one. The notation is the same as in Def. \ref{def:arrow}
\end{definition}
Again, the infimum in the above definition can be achieved,
see Sec. \ref{ach}.  We then obtain 
\begin{proposition}
\label{prop:arrowcpl}
Let  $f$ be a  function defined on states of system $X$,  which is 
subextensive and asymptotic  continuous. Then function $\farrowcpl$ is also 
asymptotically continuous. Moreover, the constant in asymptotic continuity 
condition depends only on dimension of system $X$. 
\end{proposition}
The proof is analogous to the proof of Prop. \ref{prop:arrow}.

\section{Applications}

\subsection{Measure of classical correlation  $C_\leftarrow $}

This proposition implies asymptotic continuity  of measure of classical correlation  $C_\leftarrow $ defined as follows \cite{HendersonVedral}:

\ben
C_{\leftarrow}(\rho_{AB})&=&\max_{B_{i}^{\dagger}B_{i}}S(\rho_A)-\sum_{i}p_{i}S(\rho^{i}_{A})
\een
where $B_{i}^{\dagger}B_{i}$ is a POVM performed on subsystem $B$,
$\rho^{i}_{A}=tr_B(I\otimes B_i\rho_{AB}I\otimes B_i^\dagger)/p_i$
is remaining state of $A$ after obtaining the outcome $i$ on $B$, and
$p_i=tr_{AB}(I\otimes B_i\rho_{AB}I\otimes B_i^{\dagger})$. Notice that we can rewrite $C_\leftarrow $:
\ben
C_\leftarrow(\varrho_{AB}) =\max_{B_{i}^{\dagger}B_{i}}\sum_i p_i(S(\sum_ip_i\varrho_A^i)-S(\varrho_A^i))
\een
So $C_\leftarrow $ is a kind of function build by "arrowing", where $f:\scal(\hcal_A)\to R$ 
acting on states of system $A$ if of the form:
\ben
f(\varrho_A^i)=S(\sum_ip_i\varrho_A^i)-S(\varrho_A^i)
\een
Function $f$ is asymptotically continous, 
beacuse entropy von Neumann $S$  possess this feature. So whereby of Proposition \ref{prop:arrow} quantity $C_{\leftarrow}$ is also
asymptotically cointinous.


\subsection{Intrinsic conditional information}

Consider the following function called  intrinsic conditional information: 
$I(X;Y\downarrow E)$ \cite{intI} between X and Y given E defined as 
\ben
I(X;Y\downarrow E)=\inf_{P_{\bar{E}|E}}I(X;Y|\bar{E})=
\inf_{P_{\bar{E}|E}}\sum_e p(\bar{e})I(X;Y|\bar{ E}=\bar{e})
\een
where $P_{\bar{E}|E}$ is a classical channel,  
$I(X;Y|\bar E=\bar{e})$ is mutual information between  X and Y given $\bar{ E}=\bar{e}$
and $p(\bar{e})$ 
is probability that we have outcome $\bar{e}$ on subsystem $\bar{E}$.  
The quantity $I(X;Y|\bar{E})=\sum_e p(\bar{e})I(X;Y|\bar{E}=\bar{e})$ is called conditional information. 
It is known \cite{minINT} that  infimum in definition of intrinsic conditional information is achievable. 
It is enough to take minimum over $P_{\bar{E}|E}$ with the system $\bar{E}$ of size of $E$. 

One easily finds, that that intrinsic information is  a particular case of "arrowing". 
Indeed, for a given classical channel 
$P_{\bar{E}|E}$ 
with conditional probabilities $\{p_{\bar{e}|e}\}$ we consider POVM given by Kraus operator 
$A_{\bar{e}}=\sum_e \sqrt{p_{\bar{e}|e}}|e \> \< e|$.
Now, if we embedded in natural way our distribution into set of quantum states, 
then we see that our definition \ref{def:arrow} reproduces the above quantity.
 
 If we notice that the mutual information itself is asymptotically continuous 
( it is sum of entropies, each of them being asymptotically continuous 
due to Fannes inequality (\ref{eq:Fannes})), then we will see that the asymptotic continuity of intrinsic conditional 
information follows from our theorem.



\section{Convex roof functions}
Here we present asymptotic continuity of functions  
constructed from other asymptotically continuous function $f$ 
by means of {\it convex roof} \cite{Uhlmann-roof}.
We will distinguish between 
{\it pure} and {\it mixed} convex roof. The
pure convex roof is generalization of definition 
of entanglement of formation $E_F$ given in \cite{BDSW1996}. It was proposed 
and investigated in Ref. \cite{Uhlmann-roof} and called there just convex roof.

\subsection{Pure convex roof}

\begin{definition}
For a function $f$ defined on pure states its {\it pure convex roof} $\fproof$ 
is a function defined on all states, given by 
\ben
\fproof(\varrho)=\inf_{\{p_k,\psi_k\}}\sum_k p_k f(\psi_k)
\een
where infimum is taken over all finite pure ensembles $\{p_k,\psi_k\}$, 
satisfying $\varrho = \sum  p_k|\psi _k\>\<\psi _k|$.
\end{definition}

It is useful to represent convex roof in a different way (cf. \cite{Nielsen-cont}),
to make explicit, that operation of pure convex roof is actually 
arrowing. Indeed,  for any state $\varrho $ acting 
on Hilbert space $\hcal_X$ of dimension  $d_X$ we can  construct its purification  
i.e. pure state $\varphi_\varrho$ acting on Hilbert space $\hcal_X\ot \hcal_E$ 
(with $\dim \hcal_E=\dim \hcal_X$)
such that
\ben
\tr_{\hcal_{anc}}\varphi_\varrho=\varrho
\een
Moreover for any pure decomposition of $\varrho$, given by $\{p_k,\psi_k\}$ 
there exists a complete POVM on $\hcal_{anc}$ which 
gives such ensemble on system $X$, and vice versa: any POVM gives rise to 
some pure decomposition. 

Then we can rewrite $\fproof$ as infimum over  measurements $\mathcal{M}$
\ben
\fproof(\varrho)=\inf_{\sum p_k|\psi _k\>\<\psi _k|=\varrho}
\sum_k p_k f(\psi _k)
\label{eq:roof-M}
\een
Consequently, we have 
\be
\fproof(\varrho_X)=\farrowcpl(\varphi^\varrho_{XE})
\label{eq:pure-roof-arrow}
\ee
where the equality holds for arbitrarily fixed purification $\varphi^\varrho_{XE}$ 
of the state $\varrho_X$. 
Having rewritten pure convex roof in terms of arrowed function, 
we can easily prove its asymptotic continuity, by use of the proposition 
\ref{prop:arrow}.  

\begin{proposition}
Let $f$ be a function, which is subextensive and asymptotically  continuous. 
Then its convex roof $\fproof $ is also asymptotically continuous.
\label{prop:cr}
\end{proposition}

\textbf{Proof.}\\
We will use following inequalities \cite{FuchsGraaf}:
\ben
1-F(\varrho, \sigma )\leq \frac{1}{2}||\varrho- \sigma ||_1\leq\sqrt{1-F(\varrho, \sigma )}
\label{in}
\een
where $F(\varrho, \sigma )=\sqrt{\sqrt{\varrho} \sigma \sqrt{\varrho }}$ is fidelity
\cite{Uhlmann-fidelity, Jozsa-fidelity}. 
The fidelity can be also expressed as follows
\be
F(\varrho,\sigma)=\sup|\<\psi_\varrho|\psi_\sigma\>|
\ee
where supremum is taken over all $\psi_\varrho$ and $\psi_\sigma$ 
which are purifications of states $\varrho$ and $\sigma$. 
The supremum is achievable. 

Consider now arbitrary states $\varrho$ and $\sigma$ let $\varepsilon=||\varrho-\sigma||_1$.
We want to estimate $\fproof(\varrho)-\fproof(\sigma)$. 
Since the representation (\ref{eq:pure-roof-arrow}) does not depend on the 
choice of purification, we take such purifications $\psi_\varrho$ and 
$\psi_\sigma$, that 
\be
F(\varrho,\sigma)=F(\psi_\varrho,\psi_\sigma)
\ee
Then we have 
\be
\bigl\|\, |\psi_\varrho\>\<\psi_\varrho|-|\psi_\sigma\>\<\psi_\sigma|\,\bigr\|_1 \leq 2 \sqrt{1- F(\psi_\varrho,\psi_\sigma)}
=2\sqrt{1-F(\varrho,\sigma)} \leq 2 \sqrt{||\varrho-\sigma||_1/2}=\sqrt{2\varepsilon}
\ee
Since we assume that $f$ is asymptotically continuous and subextensive,
we can use Prop. \ref{prop:arrow}
to get 
\be
|\fproof(\varrho)-\fproof(\sigma)|=
|\farrowcpl(\psi_\varrho)-\farrowcpl(\psi_\sigma)|\leq K \sqrt{2\varepsilon} \log d_X
+ O(\sqrt{2\varepsilon})
\ee
This ends the proof. \blacksquare

{\bf Remark.} Notice that however  we have here $\sqrt{2\varepsilon}$ instead of $\varepsilon $,
 but we think  that it does not change  essence of condition referring asymptotic continuity. 

\section{Mixed convex roof}
Analogously to pure convex roof we can define {\it mixed convex roof}.
\begin{definition}
Let $f$ be a function and $\varrho$ be a state then we can define 
function mixed convex roof $\fmroof$ as follows
\ben
\fmroof(\varrho)=\inf_{\{p_k,\varrho_k\}}\sum_k p_k f(\varrho  _k)
\een
where infimum is taken over all  ensembles $\{p_k,\varrho_k\}$, 
where $\varrho = \sum p_k\varrho _k$.
\end{definition}

Similarly as in the case of pure convex roof we can show that 
\be
\fmroof(\varrho_X)=\farrow(\psi^\varrho_{XE})
\ee
where, again, $\psi^\varrho_{XE}$ is arbitrarily fixed purification of $\varrho_X$. 

Therefore, with analogous proof as that of Prop. \ref{prop:cr}, we obtain

\begin{proposition}
Let $f$ be subextensive and asymptotically continuous function then 
function mixed convex roof $\fmroof$ is also asymptotically continuous.
\end{proposition}

\section{Applications}

\subsection{Pure convex roof of measure of entanglement for tripartite pure states}

Consider the quantity $E$ \cite{revers} which is equal to 
 sum of measure of entanglement for bipartite state applied for subsystem of tripartite state:
\be
E(\varrho_{ABC})=E_R(\varrho_{AB}) + S(\varrho_C) 
\ee
where S is von Neumann entropy and $\varrho_{AB}=\tr_C\varrho_{ABC}$,  $\varrho_{C}=\tr_{AB}\varrho_{ABC}$ and $E_R$ 
is relative entropy distance from set of separable states.
Now, we can consider  pure convex roof of  function $E$:

\ben
\proof{E}(\varrho_{ABC})=\inf_{\varrho_{ABC} 
=\Sigma  p_k|\psi^k\><\psi^k|_{ABC}}\sum_k p_k E(|\psi _{ABC}^k\>)
\een
Note that $E$ is subextensive and asymptotically continuous, because 
relative entropy distance and entropy possess these feature. Thus 
the Proposition \ref{prop:cr} implies that convex roof of 
this function $\proof{E}$ is also asymptotically continuous.

\subsection{Entanglement of formation} 
Proposition \ref{prop:cr}
implies asymptotic continuity 
of entanglement of formation $E_F$ (which was first shown in \cite{nielsen}) defined 
as \cite{BDSW1996}
\ben
E_F(\varrho_{AB})=\inf_{\varrho_{AB} =\Sigma  p_k|\psi _k\>\<\psi _k|}\sum_k p_k S_A(|\psi _k\>)
\een
where $S_A$ is a von Neumann entropy of subsystem A of state. In original 
definition infimum is taken over all {\it pure} ensembles, but
notice that in this case infimum over all ensembles reduce to infimum over 
pure ensembles. Thus 
\ben
E_F(\varrho_{AB})=\inf_{\varrho_{AB} =\Sigma  p_k\varrho_k}\sum_k p_k S_A(\varrho  _k)
\een
This is implied by concavity of von Neumann entropy:
\ben
\sum_kp_kS_A(\varrho_k)=\sum_kp_AS_A(\sum_iq_i^k |\varphi _i^k\>\<\varphi _i^k|)\geq
\sum_k p_k\sum_iq_i^k S_A(|\varphi _i^k\>)=\sum_{k,i} p_k q_i^k S_A(|\varphi _i^k\>)
\een
So for every mixed ensemble we can find a pure ensemble which gives 
no grater value of function $E_F$ than a mixed ensemble.

\subsection{Pure and mixed convex roof of mutual information.} 
Now, we show example of function for which pure and mixed convex 
roof are not equal to each other. 
Consider the following functions:
\ben
&&\proof{I_M}(\varrho_{AB})=\inf_{\varrho_{AB} = \Sigma  p_k|\psi _k\>\<\psi _k|}
\sum_k p_k I_M(|\psi _k\>)\\
&&\mroof{I_M}(\varrho_{AB})=\inf_{\varrho_{AB} = \Sigma  p_k\varrho _k}
\sum_k p_k I_M(\varrho  _k)
\een
where $I_M$ is mutual information 
$I_M=S_A(\varrho_{AB})+S_B(\varrho_{AB})-S(\varrho_{AB})$. 
In our terminology, the functions are pure and convex roof of 
quantum mutual information. The second one was introduced in \cite{Christandl:2002}.
Notice that for 
a pure convex roof we have 
\ben
\proof{I_M}(\varrho_{AB})=
2\inf_{\varrho_{AB} = 
\Sigma  p_k |\psi _k\> \<\psi _k|}\sum_k p_k S_A(|\psi _k\>)=2E_F(\varrho_{AB})
\een
Let $\varrho_{as}$ be antysymmetric state state:
\be
\varrho_{as}=\frac{1}{d^2-d}(I-V)
\ee
where V is a unitary flip operator V acting on Hilbert space 
$\mathcal{C}^d\otimes \mathcal{C}^d$ system defined by  
$V\phi\otimes\varphi=\varphi\otimes\phi$.
We know that \cite{VW01}
\be
E_F(\varrho_{as})=1
\ee
So $\proof{I_M}(\varrho_{as})=2$. Then we have the following inequality
\ben
\mroof{I_M}(\varrho_{as})\leq I_M(\varrho_{as})=2\log d - S(\varrho_{as})=
\log\frac{2d}{d-1}
\een
So for $d \geq 3$ we have that  $\mroof{I_M}(\varrho_{as} )\neq \proof{I_M}(\varrho_{as})$.

\section{Achieving infimum in definition of  arrowing }
\label{ach}
We prove that in definition of arrowing 
the  infimum is achievable, so that it can be replaced by minimum. First we 
prove the following lemma.
 \begin{lemma}
 Let $\{p_i\}$ be a probability distribution then
 any convex combination $ \sum_ip_ix_i$, where 
$x_i=(\varrho _i, f(\varrho _i))$, equal to  $\sum p_i(\varrho _i, f(\varrho _i))$
 can be written as a  convex combination $\sum q_i(\varrho _i, f(\varrho _i))$ consisting of $n+1$ (or less) ingredients,
 where $n$ is a dimension of space on which is acting $x_i$. So 
\ben
\sum_ip_i\varrho_i=\sum_{i=1}^{n+1}q_i\varrho_i \quad \textrm{and} \quad \sum_ip_if(\varrho_i)=\sum_{i=1}^{n+1}q_if(\varrho_i)
\een 
 \label{car}
 \end{lemma}
{\it Proof.}
Let $\tilde{f}=\sum_ip_if(\varrho_i ) $ where  $\varrho=\sum_ip_i\varrho _i$ is a state acting on Hilbert space $\mathcal{H}$.
Let $x_i=(\varrho_i,f(\varrho _i) )$ be a point from a convex set  
$\mathcal{S}=co(\varrho_i,f(\varrho_i))$. Then
\be
(\varrho ,\tilde{f})=(\sum_ip_i\varrho _i,\sum_ip_if(\varrho_i ))=\sum_ip_i (\varrho _i,f(\varrho_i )) \in \mathcal{S}
\ee  
Using Caratheodory 's Theorem we have that there exists such set of probability distribution consisting of 
$n+1$ or less elements that
\be
(\varrho ,\tilde{f})=\sum_iq_i (\varrho _i,f(\varrho_i )) 
\ee
So $\varrho =\sum_iq_i \varrho _i $ and $\tilde{f}=\sum _iq_if(\varrho_i )$. This ends the proof. \blacksquare

Now, we use above lemma to proof that infimum in function $\farrow(\varrho_{XE})$ is achievable.
Let $\psi_{AXE}$ be a purification of state $\varrho_{XE}$. Then if we make measurement $\mathcal{M}$ on subsystem $E$ we get ensemble
$\{p_i,\varrho^{AX}_i\}$ on subsystem $AX$. Lets define function $\tilde{f}$ such that for any given function $f$
 \be
 \tilde{f}(\varrho_i^{AX})=f(\varrho_i^{X})
 \ee
 where $\varrho_i^{X}=\tr_A\varrho_i^{AX}$
 Then 
\be
f_\downarrow (\varrho^{XE})=\inf_{\mathcal{M}}\sum_ip_if(\varrho_i^{X})=
\inf_{\mathcal{M}}\sum_ip_i\tilde{f}(\varrho_i^{AX})
\label{f1}
\ee
Notice that for function $\tilde{f}$ and state $\psi_{AXE} $ we can define
\be
f_\downarrow (\psi_{AXE})=\inf_{\mathcal{M}}\sum_ip_i\tilde{f}(\varrho_i^{AX})
\label{f2}
\ee
where we treat subsystem $AX$ as a one subsystem and $E$ as a second. 
Note also that 
\be
f_\downarrow (\psi_{AXE})=\inf_{\{p_i,\varrho_i^{XE}\}}\sum_ip_i\tilde{f}(\varrho_i^{AE})
\ee
because we can always find such measurement made on subsystem $E$ of  state $\psi_{AXE}$, which give us 
 ensemble $\{q_i,\varrho^{AX}_i\}$.

Then using Lemma \ref{car} we know that there exists other finite ensemble 
$\{q_i,\varrho^{AX}_i\}$ such that
\be
\sum_ip_i\varrho^{AX}_i =\sum_i^{d+1}q_i\varrho^{AX}_i \quad \textrm{and}\quad  \sum_ip_i\tilde{f}(\varrho^{AX}_i) =\sum_i^{d+1}q_i\tilde{f}(\varrho^{AX}_i) 
\ee
where $d$ is dimension of space on which is acting $\sum_ip_i\varrho^{AX}_i$. 
So for function $f_\downarrow (\psi_{AXE})$ infimum over measurement 
is effectively equal to infimum over bounded finite set of ensembles, so we have infimum over compact states. 
This implies that there exists extremal point belonging to $\mathcal{S}$, 
so  infimum for this function is achievable. 
If we are loking at formules (\ref{f1}) and (\ref{f2}) we can see that
$f_\downarrow (\psi_{AXE}) = f_\downarrow (\varrho^{XE})$, 
what implies that  for any given state $\varrho^{XE}$ function $f_\downarrow (\varrho^{XE})$ achieves infimum.

{\bf Acknowledgments.} 
We would like to thank Andreas Winter for helpful discussion on 
achievability of infima.
This work is supported by Polish Ministry of Scientic Research and Information Technology under 
the (solicited) grant no. PBZ-MIN-008/P03/2003, EU grants RESQ (IST-2001-37559), 
QUPRODIS (IST-2001-38877) and EC IP SCALA.

\section{Appendix}
Now we will present other version of Proposition \ref {pr1}. We will use Cauchy type 
conditions for asymptotic continuity and show that 
they are also equivalent to robustness under admixtures.

\begin{proposition}
Let f be a  function,
then the  following conditions are equivalent:
\ben
&&1)\forall_{\varepsilon >0} \quad \exists_{\delta>0} \quad\forall_{\varrho,\sigma }
 \quad ||\varrho-\sigma ||_1\leq \delta \Longrightarrow   |f(\varrho)-f(\sigma ) |\leq  K_1\varepsilon \log d +O(\varepsilon )\\
&&2)\forall_{\varepsilon>0} \quad \exists_{\delta>0} \quad\forall_{\varrho,\sigma } \quad |f((1-\delta ) \varrho+\delta \sigma )-f(\varrho )|
\leq K_2 \varepsilon \log d+O(\varepsilon ) \\
\een
$K_1$, $K_2$ are constants and 
$O(x)$ is  any function that satisfies (i) $O(x)$ converges to 0
when $x$ converges to 0 and (i) $O(x)$ depends only on $x$
(so in our particular case, it will not depend on dimension).

\end{proposition}

\textbf{Proof.}\\
$"1 \Rightarrow 2"$
Let $\varepsilon >0 $ be fix then there exists such $\delta >0$ that for any states $\varrho $  
and $\sigma $, the following conditions is fulfilled
\be
||\varrho-\sigma  ||_1\leq \delta \Longrightarrow   |f(\varrho)-f(\sigma ) |\leq  K_1\varepsilon \log d +O(\varepsilon )
\ee
Notice that there exists such $\delta_1=\frac{\delta}{2} $
\be
||\varrho -((1-\delta_1 ) \varrho+\delta_1 \sigma )||_1=\delta_1 ||\varrho -\sigma ||_1\leq 2\delta_1=\delta  
\ee
this implies that
\be
|f((1-\delta_1 ) \varrho+\delta_1 \sigma )-f(\varrho )|\leq K_1 \varepsilon \log d+O(\varepsilon )=K_2 \varepsilon \log d+O(\varepsilon )
\ee
$"2 \Rightarrow 1"$\\
Let $\varepsilon >0$ then there exists such $\delta >0$ that
\ben
\forall_{\varrho,\sigma } \quad |f((1-\delta ) \varrho+\delta \sigma )-f(\varrho )|
\leq K_2\varepsilon \log d+O(\varepsilon ) 
\een
Let $\varrho_1$, $\varrho_2$ be state that
\be
||\varrho_1-\varrho_2||_1=\delta_1\leq \delta 
\ee
Analogously to proof of  Theorem \ref{pr1} 
\ben
\exists_{\sigma, \gamma_1 \gamma_2} \quad\sigma =(1- \delta _1)\varrho _1+\delta _1 \gamma _1
=(1- \delta _1)\varrho _2+\delta _1 \gamma _2  
\een
\ben
&&|f(\varrho _2)-f(\varrho _1)|\leq |f(\varrho _2)-f(\sigma )|+|f(\sigma)-f(\varrho _1)|=\\
&&|f(\varrho _2)-f((1- \delta _1)\varrho _2+\delta _1 \gamma _2  )|+
|f((1-\delta _1)\varrho _1+\delta _1 \gamma _1)-f(\varrho _1)|
\leq 2K_2 \log d +2 O(\varepsilon ) =K_1 \log d + O(\varepsilon )
\een
This ends the proof.\blacksquare

\bibliography{refmich,refbasia}

\end{document}